\font\twelverm=cmr10 scaled 1200    \font\twelvei=cmmi10 scaled 1200
\font\twelvesy=cmsy10 scaled 1200   \font\twelveex=cmex10 scaled 1200
\font\twelvebf=cmbx10 scaled 1200   \font\twelvesl=cmsl10 scaled 1200
\font\twelvett=cmtt10 scaled 1200   \font\twelveit=cmti10 scaled 1200
\skewchar\twelvei='177   \skewchar\twelvesy='60
\def\twelvepoint{\normalbaselineskip=12.4pt
  \abovedisplayskip 12.4pt plus 3pt minus 9pt
  \belowdisplayskip 12.4pt plus 3pt minus 9pt
  \abovedisplayshortskip 0pt plus 3pt
  \belowdisplayshortskip 7.2pt plus 3pt minus 4pt
  \smallskipamount=3.6pt plus1.2pt minus1.2pt
  \medskipamount=7.2pt plus2.4pt minus2.4pt
  \bigskipamount=14.4pt plus4.8pt minus4.8pt
  \def\rm{\fam0\twelverm}          \def\it{\fam\itfam\twelveit}%
  \def\sl{\fam\slfam\twelvesl}     \def\bf{\fam\bffam\twelvebf}%
  \def\mit{\fam 1}                 \def\cal{\fam 2}%
  \def\tt{\twelvett}
  \textfont0=\twelverm   \scriptfont0=\tenrm   \scriptscriptfont0=\sevenrm
  \textfont1=\twelvei    \scriptfont1=\teni    \scriptscriptfont1=\seveni
  \textfont2=\twelvesy   \scriptfont2=\tensy   \scriptscriptfont2=\sevensy
  \textfont3=\twelveex   \scriptfont3=\twelveex  \scriptscriptfont3=\twelveex
  \textfont\itfam=\twelveit
  \textfont\slfam=\twelvesl
  \textfont\bffam=\twelvebf \scriptfont\bffam=\tenbf
  \scriptscriptfont\bffam=\sevenbf
  \normalbaselines\rm}



{\obeylines\gdef\
{}}
\def\singlespace{\baselineskip=\normalbaselineskip}

\def\oneandahalfspace{\baselineskip=\normalbaselineskip
  \multiply\baselineskip by 3 \divide\baselineskip by 2}
\def\doublespace{\baselineskip=\normalbaselineskip \multiply\baselineskip by 2}

\newcount\firstpageno
\firstpageno=2
\footline={\ifnum\pageno<\firstpageno{\hfil}\else
{\hfil\twelverm\folio\hfil}\fi}
\let\rawfootnote=\footnote		
\def\footnote#1#2{{\rm\singlespace\parindent=0pt\rawfootnote{$^#1$}{#2}}}

\def\dateline{\rightline{\ifcase\month\or
  January\or February\or March\or April\or May\or June\or
  July\or August\or September\or October\or November\or December\fi
  \space\number\year}}
\def\received{\vskip 3pt plus 0.2fill
 \centerline{\sl (Received\space\ifcase\month\or
  January\or February\or March\or April\or May\or June\or
  July\or August\or September\or October\or November\or December\fi
  \qquad, \number\year)}}

\input epsf
\parskip=\medskipamount
\twelvepoint		
\def\oneandahalfspace{\baselineskip=\normalbaselineskip
  \multiply\baselineskip by 3 \divide\baselineskip by 2}
\doublespace
\overfullrule=0pt	
\raggedright
\pretolerance=10000  

\line{}
\centerline{\bf SIZE OF THE VELA PULSAR'S RADIO EMISSION REGION: 500~km}
\vskip 1truein
\par\noindent{\bf C.R. Gwinn$^1$, M.J. Ojeda$^1$, M.C. Britton$^1$,
J.E. Reynolds$^2$, D.L. Jauncey$^2$, E.A. King$^{2,3}$,
P. M. McCulloch$^3$, J.E.J. Lovell$^3$,
C.S. Flanagan${^4}$, D.P. Smits${^4}$,
R.A. Preston${^5}$, \& D.L. Jones${^5}$}
\vskip 0.2 truein
\par\noindent $^1$ Physics Department, University of California, Santa
Barbara, California, 93106, USA
\par\noindent $^2$ Australia Telescope National Facility,
Epping, New South Wales, 2121, Australia
\par\noindent $^3$ Physics Department, 
University of Tasmania, Hobart, 7001, Tasmania, Australia
\par\noindent $^4$ Hartebeesthoek Radio Astronomy Observatory,
Krugersdorp, Transvaal, South Africa
\par\noindent $^5$ Jet Propulsion Laboratory, California Institute
of Technology, Pasadena, California, 91109, USA
\vfill\eject
\oneandahalfspace
\centerline{\bf ABSTRACT}

We use interstellar scattering of the Vela pulsar to determine
the size of its emission region. From 
interferometric phase variations on short baselines,
we find that radio-wave scattering broadens the source
by $3.4\pm 0.3$~mas along the major axis at 
position angle $81\pm 3$ degrees.
The ratio of minor axis to major axis is 
$0.51\pm 0.03$.
Comparison of angular and temporal broadening indicates
that the scattering material lies in the Vela supernova remnant. From 
the modulation of the pulsar's scintillation on very short
baselines,
we infer a size of 500~km for the pulsar's emission region.
We suggest that radio-wave refraction within the pulsar's
magnetosphere may plausibly explain this size.
\vfill\eject
\centerline{\bf 1. INTRODUCTION}

Although magnetized, rotating neutron stars have been observed as
pulsars for more than 25 years, the processes by which strong magnetic
fields and rapid rotation give rise to emission at radio and other
wavelengths, and allow pulsars to lose spin energy, are poorly
understood. Proposed sites for the emission range from a few meters
above the magnetic pole of the neutron star out to near the light
cylinder, where corotating fields and matter would travel at the speed
of light (e.g. Ruderman \& Sutherland 1975; Arons \& Scharlemann 1979;
Cheng, Ho, \& Ruderman 1986; Ardavan 1981).  Pulsars'
strong magnetic fields presumably mediate loss of their spin energy
through either magnetic dipole radiation or an electron-positron wind
traveling outward along open magnetic field lines.  Observational
tests of these models are difficult because the angular diameter of
the light cylinder is far smaller than the resolution limit for
Earth-based radioastronomical observations.

However, because electron density
fluctuations in the interstellar
plasma scatter radio waves,
an Earth-based observer
receives radiation from 
a range of angles, defining an effective aperture --
the ``scattering disk''.
The diameter of the scattering disk can exceed
1~AU, with consequent theoretical angular resolution of
nanoarcseconds, sufficient to resolve the light cylinder.
Observers have sought to exploit the high resolution offered by
interstellar scattering by comparing the diffraction pattern at
different points in the pulsar pulse (Backer 1975; 
Cordes, Weisberg, \& Boriakoff 1983).
Such studies rely on the hypothesis that the
apparent structure of the emission region changes over each pulse,
as the pulsar rotates.
Uncommon
episodes of very strong refraction show evidence
for a shift in the diffraction pattern over the pulse
(Wolszczan \& Cordes 1987). 
Because
the location of refracting material along the line of sight is
unknown, the corresponding shift of emission region
at the pulsar is not known.

\centerline{\bf 2. THEORETICAL BACKGROUND}

Fluctuations in the density of free electrons in the interstellar
plasma produce variations in the index of refraction for radio waves.
A pointlike source of radiation observed through such a medium
produces a random diffraction pattern in the plane of the observer.
The scattering is said to be ``strong'' if the deflection is great
enough that the random phase differences among different paths through
the medium are greater than $2\pi$.
Like that of most pulsars, the scattering of the Vela
pulsar discussed in this paper is very strong.
In strong scattering both
amplitude and phase of the electric field vary with location: the
diffraction pattern is complex.
The spatial scale of the 
diffraction pattern 
at the observer is 
$\lambda/\theta$, where $\theta$ is the angular broadening
and $\lambda$ is the observing wavelength.  
The observer sees the source scintillate
on a timescale 
$t_{ISS}=\lambda/\theta V_{\perp}$ 
as he moves with transverse velocity $V_{\perp}$
through the diffraction pattern.
The diffraction pattern also changes with observing wavelength 
because the 
phase differences among different lines of sight change,
with characteristic bandwidth of $\Delta\nu=1/2\pi\tau$,
where $\tau$ is the smearing time of a single pulse.
Observations of the diffraction pattern require 
averaging by less than $t_{ISS}$ in time
and $\Delta\nu$ in frequency. Such observations are said to be in
the speckle limit of interstellar scattering.

The Vela pulsar is the brightest pulsar at decimeter wavelengths.
Fluctuations in plasma density in the Vela-X supernova remnant
surrounding the pulsar strongly scatter its radiation (Desai et
al. 1992).  Careful studies indicate that scattering
takes place in a thin screen rather than in an extended medium
(Williamson 1972, Lee \& Jokipii 1976).  At our observing wavelength
of $\lambda=13$~cm, 
Desai et al. (1992) found that the scattering disk is
$\approx 1$~astronomical unit (AU) in diameter.
Our results below are consistent with this conclusion,
with some anisotropy. From 
comparison of angular broadening with 
temporal broadening, Desai et al. found that the scattering material
lies 4/5 the way
to the pulsar, in the Vela supernova remnant.
The scintillation timescale is $t_{ISS}=15$~sec,
and the decorrelation bandwidth is $\Delta\nu=60$~kHz.
The diffraction pattern at the
Earth has a characteristic spatial scale of about 4000~km; 
treated as a lens, 
the scattering disk
has a spatial resolution at the
pulsar of 
about 1000~km.  The scattering
disk thus resolves the 8500~km diameter of the light cylinder.

A propagator formalism relates electric field at the 
source to that at the observer 
(Bron \& Wolf 1980, Goodman 1985).
The
electric field 
in the plane of the observer $E({\bf u})$
is calculated as the integral over the electric field at each point
on the source ${\bf s}$, integrated again over 
all points ${\bf x}$ on the scattering screen,
taking into account the phase change introduced by the screen
and geometrical path length
and the decline of electric field amplitude with path length:
$$E({\bf u})=\int_{\rm screen} d{\bf x} 
{{e^{\imath 2\pi |{\bf d}|/\lambda}}\over{|{\bf d}|}}
e^{\imath\Phi({\bf x})}
\int_{\rm source} d{\bf s} 
{{e^{\imath 2\pi |{\bf r}|/\lambda}}\over{|{\bf r}|}}
E({\bf s}).\eqno{(1}$$
Here 
${\bf r}={\bf R}+({\bf x}-{\bf s})$
and
${\bf d}={\bf D}+({\bf p}-{\bf x})$,
where ${\bf R}$ is the separation of source plane and scattering screen
and ${\bf D}$ is the separation of screen and observer plane.
In the paraxial
approximation the deflections ${\bf s},{\bf x},{\bf b}$
are assumed small with respect to ${\bf R}$ and ${\bf D}$,,
and the integrals can be
recast as a double Fourier transform, with multiplication by a phase
factor including the screen and geometrical path length intervening
between the two transforms (Cornwell \& Napier 1988, Cornwell et al. 1989).
In this picture, a traditional lens works by
arranging that the screen phase precisely cancels the geometrical
phase, so that the double Fourier transform results in
an image of the source in the observer plane.
Perhaps unfortunately, the phase changes introduced by the interstellar 
scattering screen are random.

The electric field at the observer is thus a random phasor
integral over the screen. 
The phase of each contribution to that integral varies with 
the observer's position, in a plane at constant distance from the screen.
The phase structure function $D_{\phi}({\bf b})$ 
is the mean square variation in phase of those contributions,
for lateral separation ${\bf b}$ (e.g. Rickett 1977):
$D_{\phi}({\bf b})=\left<(\phi({\bf u}+{\bf b})-\phi({\bf u}))^2\right>.$
Here the triangular brackets $<..>$ denote
an average over an ensemble of possible scattering screens
as well as over points on individual screens that contribute to 
the phasor integral for the electric field.
The correlated flux density $C_{AB}({\bf b})$ 
for an interferometer
with baseline ${\bf b}$ is
the product of their electric fields.
Half of the difference between the 2 phasor integrals will
result in differences in electric-field amplitude
and half in differences in phase.
The phase difference between electric fields
equals the phase of $C_{AB}$,
and can be used to infer the phase structure function (Desai et al. 1992).
For short baselines,
$$D_{\phi}({\bf b})={{\left(<{\rm Im}\{C_{AB}({\bf b})\}>\right)^2}\over
{<({\rm Re}\{C_{AB}({\bf b})\})^2>}}.\eqno{(2}$$
Thus, measurements of the phase of the correlated flux
density, on short baselines, yield the phase structure function.
Because the phase structure function varies 
with baseline $b$ nearly as $b^2$
this measurement yields the size of the scattering disk $\theta_H$:
$$D_{\phi}({\bf b})=
\left[{{\pi}\over{\sqrt{2\ln 2}}}{{\theta_H b}\over{\lambda}}\right]^2
\eqno{(3}$$
Here $\theta_H$ is the angular full width at half maximum intensity
of the scattering disk.
Slightly more complicated expressions describe the more
general situation where the scattering disk is elongated
(Gwinn et al. 1988).

Interstellar scattering acts like an imperfect optical system
in that the diffraction pattern in the plane of the observer
is the convolution of the response to a point source
(the ``point spread function'') with
an image of the source.
In principle, an diffraction-limited image of the source 
and the phase variations of the scattering screen
can both be extracted from the diffraction pattern
(Cornwell et al. 1989,
Narayan \& Cornwell 1993).
On a less ambitious scale, 
the modulation of scintillation provides a measure of
source size
(most simply, stars twinkle, planets do not).
The modulation index $m$ provides a quantitative measure
of the source size.
The modulation index is the rms fractional variation of intensity:
$m^2=\langle I^2-\langle I\rangle ^2\rangle /\langle I\rangle ^2$.
For a point source $m=1$.
The modulation index is most easily analyzed in the 
domain of wavenumber, the Fourier conjugate of position in
the observer plane.
The spectrum of intensity fluctuations for a point source
in this domain is (Rickett 1977, Codona et al. 1986):
$$P_{0}({\bf q})=
\left|\int d{\bf u}\quad
e^{\imath ({\bf q}\cdot{\bf u})}I({\bf u})\right|^2
=\int d{\bf b}\quad
e^{\imath ({\bf q}\cdot{\bf u})} e^{-D_{\phi}({\bf b})}.\eqno{(4}$$
The integrals are over all values of ${\bf u}$ and ${\bf b}$.
The modulation index is
(Salpeter 1967; Cohen, Gundermann, \& Harris 1967):
$$m^2={{1}\over{4\pi^2 <I>^2}}
\int d{\bf q}\quad
P_{0}({\bf q})\quad
\left|V\left({{\lambda}\over{2 \pi}}R M {\bf q}\right)\right|^2,\eqno{(5}$$
where $V({\bf b})$ is the traditional interferometric
visibility as a function of baseline ${\bf b}$.
Here 
$M=R/D$, where 
$D$ is the distance from source to scattering screen,
and $R$ is the distance from scattering screen to observer;
$M$ plays a role analogous to the magnification of a lens.
Note that the baselines sampled in the expression
are of order the diameter of the scattering disk, or a few
AU for the Vela pulsar.
For a Gaussian souce with full width at half maximum 
$s_x$ and $s_y$ in the x- and y-directions,
the modulation index is 
$$m^2=\left[
\left( 1+\left({{\pi}\over{2\ln 2}}{M s_x \theta_Hx}\right)^2\right)
\left( 1+\left({{\pi}\over{2\ln 2}}{M s_y \theta_Hy}\right)^2\right)
\right]^{-1/2}.\eqno{(6}$$

The square of the modulation index is 
the second moment of the
distribution function of intensity,
$P(I)$.
For a point source in strong scattering,
the random phasor integrals produce a Gaussian probability distribution
function for electric field, 
and an exponential distribution for its square, the intensity
(Scheuer 1968):
$P(I)=\exp(-I/I_0)/I_0$.
Here the  exponential scale $I_0$ is the average intensity.
For a source small compared with the 
nominal resolution of the scattering
disk, 
variations of the phasor integral introduce
2 additional sharply-declining exponentials,
corresponding to variations of the phasor integral along
the 2 dimensions of the source.
These additional exponentials have respective scales 
$(\pi M \theta_{Hx} s_x/4\ln 2\lambda)^2 I_0$ and 
$(\pi M \theta_{Hy} s_y/4\ln 2\lambda)^2 I_0$.
The net distribution is the original distribution with the
difference of the 2 sharply-declining exponentials subtracted from it.
The relative normalizations 
of these 3 exponentials are set by the requirements that
at $I=0$,
$P(I)=0$ and $dP/dI=0$.
\vfill\eject
\centerline{\bf 3. OBSERVATIONS AND ANALYSIS}

We observed the diffraction pattern of the Vela pulsar interferometrically
in October-November 1992.  For this southern-sky object we used antennas
at Tidbinilla (70 m diameter), Parkes (64 m), and Hobart (25 m) in
Australia, Hartebeesthoek (25 m) in South Africa, and the 7 antennas of
the Very Long Baseline Array of the US National Radio Astronomy Observatory
\footnote*{The National Radio  Astronomy Observatory is
operated by  Associated Universities Inc., under a  cooperative agreement with
the National Science Foundation.}
that could usefully observe the
source.  We analyzed the data recorded at each antenna
at the Haystack very-long
baseline interferometry (VLBI) correlator, forming
both autocorrelation spectra to sample the diffraction pattern at
individual antennas, and cross-power spectra to measure the relative
phase and spatial coherence between antennas.  To study possible
variation in the structure of the pulsar across the pulse we made all
measurements in 3 different gates, each covering a different range of
phases of the pulse profile.

A tremendous number of observables can be calculated from the complex
diffraction pattern.  
In this paper we concentrate on 2: the mean square
variation of phase due to scintillation; and the
the modulation index.
Phase variation on short baselines yields the 
size, elongation, and position angle of the scattering disk.
With this information,
the modulation index yields the size of the pulsar's emission region.
We will discuss results from the long baselines,
which sample the 2-dimensional structure of the 
source and its variations with pulse phase,
in another paper.

We use observations over a range of orientations of the
Tidbinbilla-Hobart baseline to determine
the phase structure function and so the size and
shape of the scattering disk.
This baseline has projected length between 500 and 850~km,
so it is much shorter than the scale of the diffraction pattern,
of 4000~km.
Figure 1 shows a comparison of our measured phase structure
function with the best-fitting models for circular
and elliptical Gaussian distributions
for the scattering disk.
The elliptical model clearly fits better.
This model has full width at half maximum of
$3.4\pm 0.3$~mas along a major axis at position angle 
$81\pm 3$ degrees, and a ratio of minor axis to major axis
of $0.51\pm 0.03$.

We find the modulation index from the distribution of intensity
on the Parkes-Tidbinbilla baseline.
This baseline is only 275~km long, so short that 
the diffraction pattern is nearly identical at the 2 antennas.
Interferometric observations greatly reduce 
effects of receiver noise and interference.
Figure 2 shows the distribution of intensity, observed
over a period of
13 minutes starting at 23:40~UT on 1992 Oct 31.
The data in the lower panel are from a single pulse gate,
comprising the 1.16~msec immediately preceding the peak of the pulse;
those in the upper panel are from a gate of the same width,
but containing
no pulsed emission.

Figure 2 also shows best-fitting models to the data.
For the ``empty'' gate this
model is Gaussian noise. The fit parameters are the strength of
noise and the normalization.
For the gate with pulsed emission,
we show the expected distribution for
a point source: this is the convolution of Gaussian
noise with an exponential, with exponential scale 
equal to that
of the observed distribution at large amplitudes.
We also show the expected distribution for a pulsar
emission region of finite size,
with full width at half maximum of 500~km.
Clearly, the model for the resolved source fits better.
Both models are adjusted to have the normalization of
the observed distribution. The size of the
emission region is the only free parameter in this plot.
The modulation index for this
best-fitting model, after removing effects of noise, is
$m=0.90$.

Averaging over several scintillation times
or bandwidths can reduce the modulation index,
but our averaging time and bandwidth are well within
the speckle limit.
Low-level emission from a non-scintillating source 
would reduce scintillation, but we detect no such source
in gates outside the pulse.
Receiver noise could reduce it,
but is well-determined from the empty gate for the low-intensity points
that show the effect.
The pulsar signal itself will increase the noise level
at times
and frequencies where the pulsar is strongest,
but not where it is weakest, as observed.
Errors in the Van Vleck correction for 1-bit recording
as implemented in the correlator software 
should likewise be greatest
for large amplitudes.

Our model assumes a circular Gaussian distribution of emission.
The emission region could be larger if the extension
is closely aligned with the nearly 
north-south minor axis of the pulsar,
up to 1500~km. 
For such large sizes,
the source would have to be highly elongated:
for a size of 1000~km along the major axis
of the scattering disk the elongation
must exceed 5:1. 
We will discuss data from the long baselines,
which probe the aspect ratio of the source, elsewhere.

\par\noindent{\bf 5. DISCUSSION}

The 2:1 elongation of the Vela pulsar's scattering disk is not unusual
for radio-wave scattering in the interstellar plasma 
(Wilkinson, Narayan, \& Spencer 1994,
Frail et al. 1994, Desai et al. 1994, Molnar et al. 1995).
Such anisotropic scattering is expected in turbulent plasmas
with strong magnetic fields
(Higdon 1984, Goldreich \& Sridhar 1995).
Theory predicts that the magnetic field should
lie perpendicular to the major axis of the scattering screen.
As inferred from polarization of synchrotron radiation,
the magnetic field in the Vela-X supernova remnant lies at
a position angle of $45^{\circ}$ at the line of sight to the
pulsar, but runs nearly north-south close by (Milne 1980).
Of course, this synchrotron emission may not emanate from the
scattering screen.

The size of 500~km is much smaller than the radius of the light
cylinder but is larger than many hypothesized emission regions,
such as the popular polar-cap models.
The extent of the emission region is far larger than that
expected for the $1/\gamma$ opening angle of emission
for electron and positron curvature radiation
along dipole magnetic field lines,,
for Lorentz factor $\gamma > 100$.
If the emission region lies closer to the pulsar than
35\% of the distance to the light cylinder,
then its 500~km size exceeds the 5\%
duty cycle of the pulsar, and
radiation from this surface would
have to be beamed into a cone narrower than the size
of the emission region as seen from the pulsar.
In the context of the Radhakrishnan \& Cooke (1969) model,
direction of the linearly-polarized pulsar emission reflects
the magnetic field direction at the emission surface.
Finite size of the emission region would lead
to depolarization,
which would be greater than that observed
even if the emission region lies
at the radius of the light cylinder.

Krishnamohan \& Downs (1983) 
made an extensive study of the pulse of the Vela pulsar
and dissected it into 4 components with different 
longitudes and polarization properties.
These components contribute with varying strength, depending
on the peak intensities of the pulse.
Their model 
assumes that radiation from one component 
arises at a single longitude of the pulsar's magnetic field
at each instant, and so is somewhat complementary to the
measurements of modulation index presented here. From 
relative longitudes and polarization
sweep rates of the different components they infer 
altitudes and longitudes 
of the different components, with a spread of
about 400~km in altitude.
It is interesting that their model yields a size of order that 
we infer from modulation index.

We suggest that the observed
size might result from magnetospheric refraction.  Magnetospheres of
pulsars must contain some plasma, to cancel the electric fields
induced by the strong rotating magnetic field
(Goldreich \& Julian 1969).
Stong winds believed to carry spindown energy away from 
the pulsar may require much greater densities 
(Ruderman \& Sutherland 1975; Arons \& Scharlemann 1979).
Refraction and scattering by this plasma can 
heavily modify radio waves propagating outward through the
plasma (Melrose \& Stoneham 1977, Arons \& Barnard 1986);
indeed, they can displace them while approximately maintaining
their original direction and polarization (Barnard \& Arons 1986),
so that a source of significant size can still
produce a relatively collimated beam.
A strong wind carrying spindown energy outward through the magnetosphere
can refract radiation of $\lambda=13$~cm wavelength 
out to a large fraction of the light-cylinder radius if
in the observer
frame if the density is $10^3-10^5\times$ the minimum required
to cancel the vacuum electric field (Goldreich \& Julian 1969)
and the typical Lorentz factor is $\gamma\approx 100-1000$.
Such parameters are expected for secondary electron-positron pairs
produced by the $\gamma\approx 10^6$ primary pairs
that would be accelerated by a polar cap.

Clearly, 
speckle VLBI can reveal much about radio emission from pulsars.
Observations on long baselines,
to be discussed elsewhere, probe the 2-dimensional
structure of the pulsar.
We also plan 
polarimetric VLBI observations of the Vela pulsar,
using the VSOP space-based VLBI antenna
in conjunction with ground radio telescopes.
Observations of other pulsars,
particularly measurements of
modulation index for pulsars 
with scattering material concentrated
close to the pulsar, such as the Crab pulsar,
could provide a measure of the size of
the emission region for other pulsars.

We thank J. Arons, J. Barnard, R. Capallo, W. Coles, R. Narayan
and B. Rickett
for useful suggestions.
This research was supported in part by 
the National Science Foundation.
The Australia Telescope is operated as a national facility
by CSIRO. The Deep Space Network is operated by the Jet Propulsion
Laboratory, California Institute of Technology, under contract with
the National Aeronautics and Space Administration.

\vfill\eject
{
\parindent=-0.25truein

%
\normalbaselineskip=8pt plus0pt minus0pt  \parskip 0pt

\centerline {\bf REFERENCES}
\smallskip

\item{} Ardavan, H. 1981, Nature, 289, 44 
\item{} Arons, J., \& Barnard, J.J. 1986, ApJ, 302, 120 
\item{} Arons, J., \& Scharlemann, E.T. 1979, ApJ, 231, 854
\item{} Backer, D.C. 1975, A\&A, 43, 395 
\item{} Barnard, J.J., \& Arons, J. 1986, ApJ, 302, 138 
\item{} Born, M., \& Wolf, E. 1989, Principles of Optics, Oxford: Pergamon
\item{} Cheng, K.S., Ho, C., \& Ruderman, M. 1986, ApJ, 300, 500
\item{} Codona, J.L., Creamer, D.B., Flatt\'e, S.M., Frehlich, R.G., \& Henyey, F.S. 1986, Radio Science, 21, 805
\item{} Cohen, M.H., Gundermann, E.J., \& Harris, D.E. 1967, ApJ, 150, 767
\item{} Cordes, J.M., Weisberg, J.M., \& Boriakoff, V. 1983, ApJ, 268, 370
\item{} Cornwell, T.J., \& Napier, P.J. 1988, Radio Sci, 23, 739-748 
\item{} Cornwell, T.J., Anantharamaiah, K.R., \& Narayan, R. 1989, JOSA,  A 6, 977
\item{} Daugherty, J.K., Harding, A.K. 1994, ApJ, 429, 325 
\item{} Desai, K.M., Gwinn, C.R., Reynolds, J.R., King, E.A., Jauncey, D., Flanagan, C., Nicolson, G., Preston, R.A., \& Jones, D.L. 1992, ApJ, 393, L75 
\item{} Desai, K.M., Gwinn, C.R., \& Diamond, P.J. 1994, Nature, 372, 754 
\item{} Frail, D.A., Diamond, P.J., Cordes, J.M. \& van Langevelde, H.J. 1994, ApJ, 427, L44 %
\item{} Goldreich, P., \& Julian, W.H. 1969, ApJ, 157, 869 
\item{} Goodman, J.W. 1985, Statistical Optics (New York: Wiley)
\item{} Gwinn, C.R., Cordes, J.M., Bartel, N., Wolszczan, A., \& Mutel, R.L. 1988, ApJ, 334, L13
\item{} Higdon, J.C. 1984, ApJ, 285, 109
\item{} Krishnamohan, S. \& Downs, G.S. 1983, ApJ, 265, 372
\item{} Lee, L.C., \& Jokipii, J.R., 1976, ApJ, 206, 735 
\item{} Melrose, D.B. and Stoneham, R.J., 1977, Proc. Astr. Soc. Australia, 3, 120
\item{} Milne, D.K., 1980, Ast Ap, 81, 293 
\item{} Molnar L.A., Mutel R.L., Reid M.J., Johnston, K.J. 1995, ApJ, 438, 708
\item{} Narayan, R., \& Cornwell, T.J. 1993, ApJ, 408, L68 
\item{} Radhakrishnan, V., \& Cooke, D.J. 1969, ApLett, 3, 225 
\item{} Ruderman, M.A., \& Sutherland, P.G. 1975, ApJ, 196, 51 
\item{} Rickett, B.J., 1977, ARA\&A, 15, 479
\item{} Salpeter, E.E., 1967, ApJ, 147, 433 
\item{} Scheuer, P.A.G. 1968, Nature, 218, 920
\item{} Wilkinson, P.N., Narayan, R. \& Spencer, R.E., 1994, MNRAS, 269, 67 
\item{} Williamson, I.P., 1972, MNRAS 157 55
\item{} Wolszczan, A. \& Cordes, J.M. 1987, 320, L35
}
\vfill\eject
\singlespace
\parindent=0.25truein
\centerline{\epsfxsize=6.20truein \epsfbox{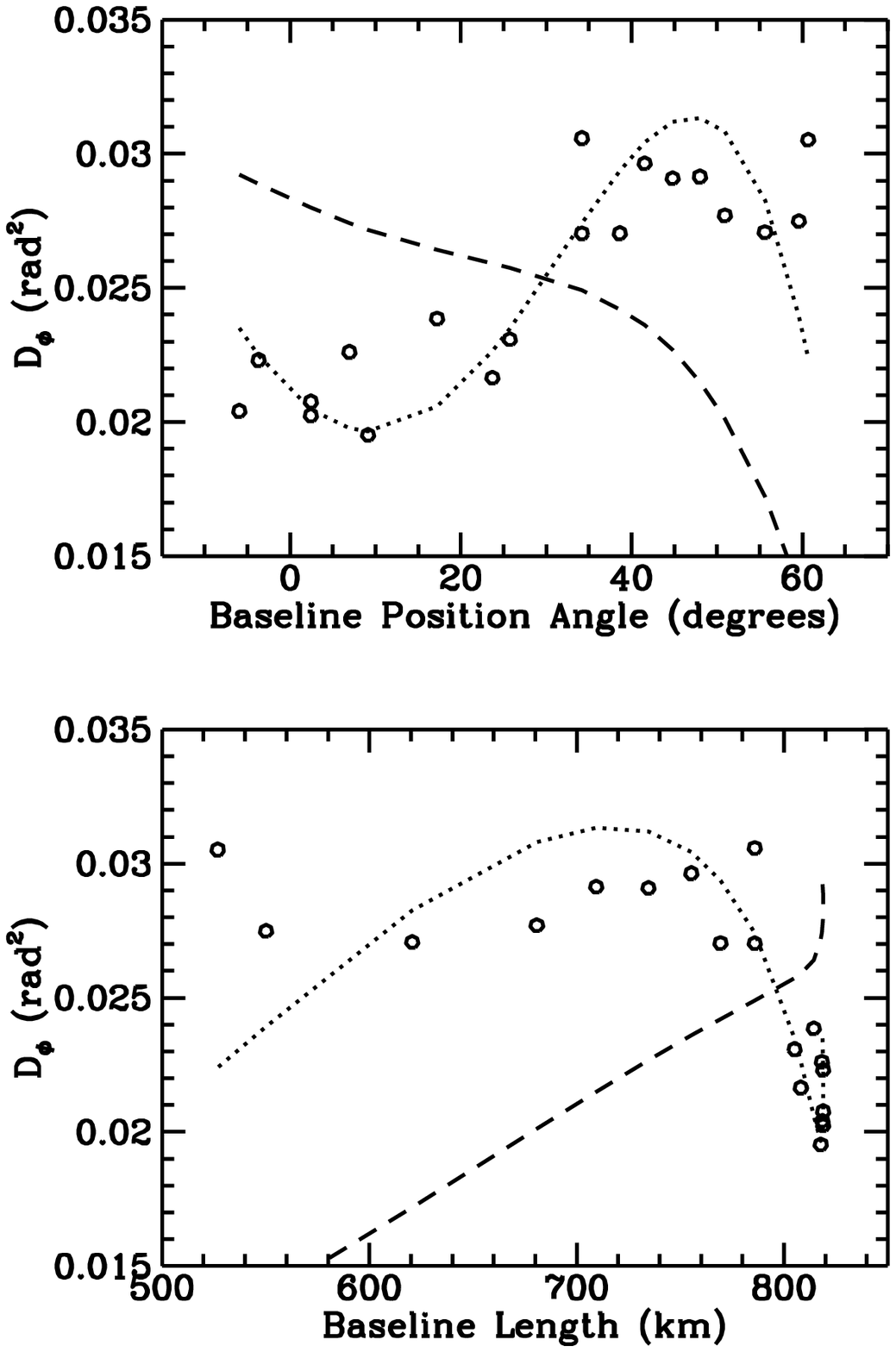}}
\vskip .13 in
\par\noindent
{\bf Fig 1:} 
Measured phase structure function of the Vela pulsar
compared with best-fitting models.
Circles show phase structure function,
as measured from real and imaginary components
of the correlated flux density,
on the Tidbinbilla-Hobart baseline.
Upper panels shows measurements plotted with
position angle of baseline, and lower panel
plotted with length of baseline.
Dashed curves show the best-fitting circular Gaussian model for the
scattering disk.
Dotted curves show the best-fitting model for an elliptical Gaussian
distribution of intensity on the sky.
The parameters of this elliptical Gaussian are:
full width at half maximum of major axis:
$3.4\pm 0.3$~mas;
position angle of major axis $81\pm 3$~deg;
ratio of minor to major axis $0.51\pm 0.026$.
The scattering material lies in the vela-X supernova
remnant surrounding the pulsar
(Desai et al. 1992);
the elongation of the scattering disk is likely due to the magnetic field
of the remnant.
\vfill\eject

\centerline{\epsfxsize=4.50truein \epsfbox{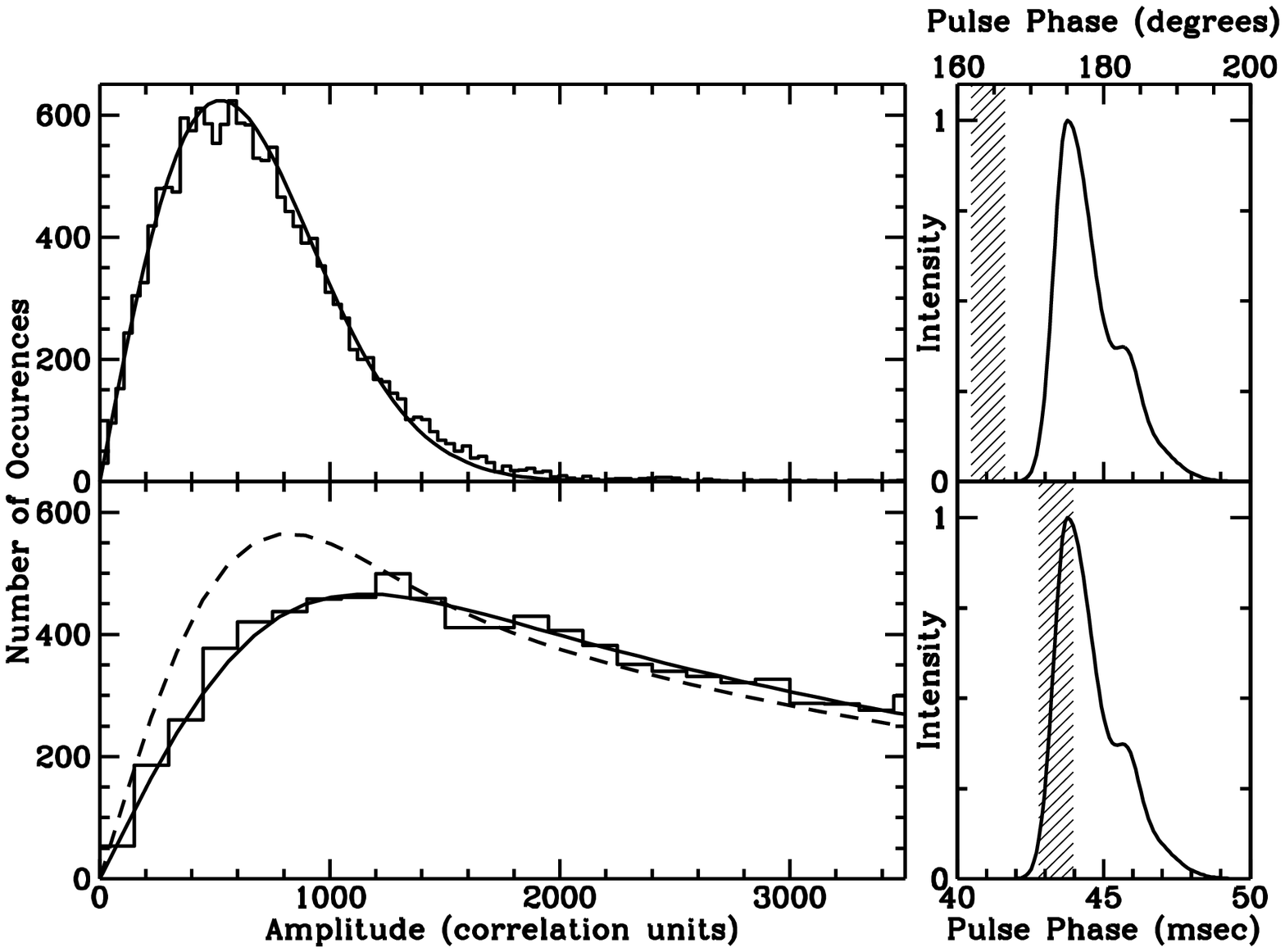}}
\vskip .13 in
{\bf Fig 2:} Distribution in amplitude of correlation function
for the Vela pulsar and for noise, on the Tidbinbilla-Parkes
baseline.
Because this baseline is much shorter than the
scale of the diffraction pattern, 
the amplitude is equivalent to intensity at
a single location.
The 13~min observation covered 4$\times$2-MHz bands between
2.275 and 2.299~GHz, starting
at 23:40~UT on 31~Oct 1992.
The amplitude was sampled with 10-sec averaging time
and frequency resolution 25~kHz: well within the
speckle limit for this pulsar.
{\it Upper:} Distribution of amplitude in a 1.16-msec gate
synchronized with the Vela pulsar,
but offset from the pulse, so that we expect zero emission.
Curve shows best-fitting distribution for Gaussian noise,
as expected in this empty gate.
The fit includes parameters for the strength of noise
and normalization.
{\it Lower:} Distribution of amplitude for a gate
including the first 1.16~msec of the pulsar pulse,
up to the time of peak emission.
At large amplitude, the distribution function follows 
the expected exponential form, which continues beyond the plot.
Dashed curve shows the purely-exponential form expected for a point source,
convolved with noise as calculated from the upper panel.
Solid curve shows the 
form expected for a
circular Gaussian emission region with full width at half maximum
of $s=500$~km. 
The size $s$ is the only free parameter in the lower panel:
the exponential scale is set by the form of the histogram at large amplitude,
and the models are normalized to have the area of the data histogram.
\bye